\documentclass[twocolumn]{el-author}
\usepackage{amsmath,amsthm}

\newcommand{\ua}{\uparrow}
\newcommand{\nc}{\newcommand}
\nc{\da}{\downarrow} \nc{\hc}{\hat{c}} \nc{\hS}{\hat{S}}
\nc{\bra}{\langle} \nc{\ket}{\rangle} \nc{\eq}{equation (\ref}
\nc{\h}{\hat} \nc{\hT}{\h{T}}\nc{\be}{\begin{eqnarray}}
\nc{\ee}{\end{eqnarray}}\nc{\rd}{\textrm{d}}\nc{\e}{eqnarray}\nc{\hR}{\hat{R}}\nc{\Tr}{\mathrm{Tr}}
\nc{\tS}{\tilde{S}}\nc{\tr}{\mathrm{tr}}\nc{\8}{\infty}\nc{\lgs}{\bra\ua,\phi|}\nc{\rgs}{|\ua,\phi\ket}
\nc{\hU}{\hat{U}}\nc{\lfs}{\bra\phi|}\nc{\rfs}{|\phi\ket}\nc{\hZ}{\hat{Z}}\nc{\hd}{\hat{d}}\nc{\mD}{\mathcal{D}}
\nc{\bd}{\bar{d}}\nc{\bc}{\bar{c}}\nc{\mc}{\mathcal}\nc{\ea}{eqnarray}\nc{\mG}{\mathcal{G}}\nc{\bce}{\begin{center}}
\nc{\ece}{\end{center}}
\date{12th December 2011}

\begin{document}

\title{An algorithm of frequency estimation for multi-channel coprime sampling}

\author{Shan Huang, Haijian Zhang, Hong Sun and Lei Yu}

\abstract{In some applications of frequency estimation, it is challenging to sample at as high as the Nyquist rate due to hardware limitations. An effective solution is to use multiple sub-Nyquist channels with coprime undersampling ratios to jointly sample. In this paper, an algorithm suitable for any number of channels is proposed, which is based on subspace techniques. Numerical simulations show that the proposed algorithm has high accuracy and good robustness.}

\maketitle

\section{Introduction}

Frequency estimation of multiple sinusoids has wide applications in communications, audio, medical instrumentation and electric systems. The methods for frequency estimation cover classical modified DFT \cite{belega2008frequency}, subspace techniques such as MUSIC \cite{schmidt1986multiple} and ESPRIT \cite{roy1989esprit} and other advanced spectral estimation approaches \cite{leung2000modified}. In general, the sampling rate of a signal is required to be higher than twice the maximum frequency component (i.e. the Nyquist rate). However, it is challenging to build sampling hardware when signal bandwidth is large. When a signal is sampled below the Nyquist rate, it often leads to aliasing and frequency ambiguity.

A number of methods have been proposed to estimate the frequencies with sub-Nyquist sampling. Zoltowski proposed a time delay method which requires the time delay difference of the two undersampled channels not greater than the Nyquist sampling interval \cite{zoltowski1994real}. By introducing properly chosen delay lines and sparse linear prediction, the method in \cite{tufts1995digital} provided unambiguous frequency estimates with low A/D conversion rates. The authors of \cite{zhou1997multiple} made use of Chinese Remainder Theorem (CRT) to uniquely determine the frequency. Based on emerging compressed sensing theory, sub-Nyquist wideband sensing algorithms and corresponding hardware were designed to estimate the power spectrum of a wideband signal \cite{mishali2011xampling}. However, these methods usually require much hardware or complex calculations, which makes the practicability discounted. In \cite{Vaidyanathan2011Sparse} and \cite{liu2015remarks}, two channels with coprime undersampling ratios are utilized to estimate the frequencies of multiple sinusoids. This method is quite simple, but in some cases, two coprime channels can not uniquely determine the estimated frequencies \cite{huang2017frequency}.

In this letter, we propose an algorithm that is suitable for any number of channels, and then utilize three coprime channels to yield accurate frequency estimates.

\section{Problem model}
Consider a signal $\bm x(t)$ containing $K$ frequency components with unknown constant amplitudes and phases
\begin{equation}\label{eq1}
 x(t) = \sum\limits_{k = 1}^K {{s_k}{e^{j{\omega _k}t}}}+ w(t) ,
\end{equation}
where $\omega _k$ and $s_k$ are the $k$-th normalized angular frequency and its corresponding complex amplitude, respectively, and $w(t)$ is zero-mean complex white Gaussian noise. When $t=1,2,\cdots,M$, it implies normal sampling, which is studied in conventional methods such as the ESPRIT algorithm \cite{roy1989esprit}. Assume that the upper limit of the frequencies $F_{H}$ is known, but we only have low-rate analog-to-digital converters whose sampling rates are much lower than the Nyquist rate.
\begin{figure}[!htbp]
\centering
\includegraphics[scale=0.6]{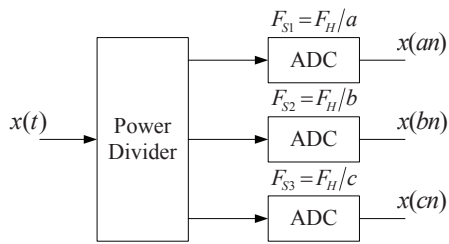}
\caption{ The block diagram of three-channel coprime sampling } \label{fig1}
\end{figure}

Our sampling strategy is sampling at three rates $F_{S1}=F_{H}/a$, $F_{S2}=F_{H}/b$ and $F_{S3}=F_{H}/c$, where $a,b,c$ are pairwise coprime integers. As shown in Fig.~\ref{fig1}, the original signal is divided into three channels, $n=1,2,\cdots$. At the same time point, three analog-to-digital converters (ADCs) start to sample at different rates. Consequently, we obtain the samples with indices
\begin{equation}\label{eq2}
 \mathcal{T }= \left\{ {a,2a, \cdots } \right\} \cup \left\{ {b,2b, \cdots } \right\} \cup \left\{ {c,2c, \cdots } \right\}.
\end{equation}

The authors of \cite{Vaidyanathan2011Sparse} use two coprime channels to increase the number of frequency components that can be estimated. However, in some cases, the frequencies can not be uniquely determined by two coprime channels. In \cite{huang2017frequency}, three coprime channels are verified to be able to resolve the frequency ambiguity in general. The following algorithm is suitable for any number of channels, but three coprime channels are recommended.
\section{Estimation algorithm}
The core of the proposed algorithm is to generate the estimate of the autocorrelation matrix from non-equidistance samples. Usually the data matrix is filled with equally spaced samples. Assuming that the samples are sampled from $t = 0$ and each $M$ samples form a column vector, we expect to construct a continuous measurement data matrix
\begin{equation}\label{eq3}
 {\bm{X}} = \left[ {\begin{array}{*{20}{c}}
{x(1)}&{x(2)}& \cdots &{x(L)}\\
{x(2)}&{x(3)}& \cdots &{x(L + 1)}\\
 \vdots & \vdots & \ddots & \vdots \\
{x(M)}&{x(M + 1)}& \cdots &{x(M + L - 1)}
\end{array}} \right],
\end{equation}
where $L$ is the number of snapshots (or measurement vectors). However, some samples are missing in this data matrix. We put the available samples into this matrix, and the positions of the missing samples are set to 0. For example, if $a,b,c=3,4,5$, $M=5$ and $L=6$, the constructed data matrix is
\begin{equation}\label{eq4}
 {\bm{X}} = \left[ {\begin{array}{*{20}{c}}
0&0&{x(3)}&{x(4)}&{x(5)}&{x(6)}\\
0&{x(3)}&{x(4)}&{x(5)}&{x(6)}&0\\
{x(3)}&{x(4)}&{x(5)}&{x(6)}&0&{x(8)}\\
{x(4)}&{x(5)}&{x(6)}&0&{x(8)}&{x(9)}\\
{x(5)}&{x(6)}&0&{x(8)}&{x(9)}&{x(10)}
\end{array}} \right].
\end{equation}
From the data matrix we calculate
\begin{equation}\label{eq5}
  {\bm{Q}} = {\bm{X}}{{\bm{X}}^\mathrm{H}},
\end{equation}
where $(*)^\mathrm{H}$ denotes the Hermite transpose operation. Note that the cumulative number of products in each position of $\bm Q$ is different. We introduce a matrix to mark the positions of the available elements in the data matrix $\bm X$, which we call the position matrix. This position matrix $\bm G$ is the same size as the matrix $\bm X$. If there is a sample available in $\bm X$, then the element in the corresponding position in $\bm G$ is set to 1, otherwise it is set to 0. For example, the position matrix for the data matrix in (\ref{eq4}) is
\begin{equation}\label{eq6}
  \bm G = \left[ {\begin{array}{*{20}{c}}
0&0&1&1&1&1\\
0&1&1&1&1&0\\
1&1&1&1&0&1\\
1&1&1&0&1&1\\
1&1&0&1&1&1
\end{array}} \right].
\end{equation}
Similarly, we can obtain
\begin{equation}\label{eq7}
  {\bm{P}} = {\bm{G}}{{\bm{G}}^\mathrm{H}}.
\end{equation}
The elements in $\bm{P}$ represent the cumulative number of products for each position in $\bm{Q}$. For two matrices of the same dimensions, the Hadamard division (or the entrywise quotient) produces another matrix where each element is the quotient of elements of the original two matrices. Denoting the Hadamard division of two matrices '$\oslash$', the estimate of the autocorrelation matrix is
\begin{equation}\label{eq8}
\bm R=\bm Q\oslash \bm P,
\end{equation}

Next, the subspace techniques can be applied to achieve high-resolution frequency estimation, such as MUSIC and ESPRIT. The above algorithm requires that each element of the matrix $\bm P$ is nonzero. The property in \cite{Vaidyanathan2011Sparse} is useful:
\newtheorem{thm1}{Theorem}
\begin{thm1}
Let $a$ and $b$ be two coprime positive integers. Given an arbitrary integer $m$ in the range $0\leq m\leq ab-1$, we can always find $n_{1}$ in the range $0\leq n_1\leq 2b-1$ and $n_{2}$ in the range $0\leq n_2\leq a-1$, such as $m=a n_1-b n_2$.
\end{thm1}
In fact, Theorem 1 can be easily proven through B\'{e}zout's identity. Even if there are only two coprime channels, we can obtain the following conclusion:
\newtheorem{thm2}[thm1]{Theorem}
\begin{thm2}
Assuming that two channels are used to sample at coprime undersampling ratios $a$ and $b$, if the number of snapshots $L$ in (\ref{eq3}) satisfies $L\geq ab$, then all the elements of the matrix $\bm P$ are nonzero.
\end{thm2}
{\bf Proof}:
When two coprime channels are used, let $a<b$, the set of the indices for the samples in ascending order is
\begin{equation}\label{eq9}
   \mathcal{T }= \left\{ a,b, \cdots, ab, ab+a, ab+b,\cdots \right\}.
\end{equation}
The measurement vectors in the data matrix $\bm X$ are formed by sliding the elements in the set $\mathcal{T }$ with the window size $M$. According to Theorem 1, an arbitrary integer $M$ satisfying $1\leq M\leq ab$ can be derived from the differences of the elements in the set
\begin{equation}\label{eq10}
  \mathcal{T}_{0} = \left\{ a,b, \cdots, ab, ab+a, ab+b,\cdots, 2ab-b, 2ab-a  \right\}.
\end{equation}
In other words,
\begin{equation}\label{eq11}
  \{1,2,\cdots,ab\}\subset\{n_{1}-n_{2}|n_{1}>n_2,~n_{1}\in \mathcal{T}_{0},~n_{2}\in \mathcal{T}_{0}\}.
\end{equation}
Note that the indices in $\mathcal{T }$ are repeated with a period of $ab$, then for an arbitrary integer $M$, we can always find two integers in $\mathcal{T }$ so that the difference between the latter and the former is $M$. The structure of the measurement vector in the window is also cycled with a period of $ab$, that is, when $L\geq ab$, all structural types can be traversed. Thus, when $L\geq ab$, all the elements of the matrix $\bm P$ are nonzero.

According to Theorem 2, when using three coprime channels with undersampling ratios of $a$, $b$ and $c$, if $L\geq min(ab,bc,ac)$, all the elements of the matrix $\bm P$ can be guaranteed to be nonzero.
\section{Simulation result}
In this section, we simulate frequency estimation of multiple sinusoids buried in noise. The signals contain $K$ frequency components with random amplitudes in the interval $[0.5,1]$ and random phase angles in the interval $[0,2\pi)$. The normalized frequencies are assumed to distribute uniformly in $(0,1]$ and their intervals are set to be larger than 0.01. Complex white Gaussian noise is added to the measurements. The coprime undersampling ratios are set to $a=3$, $b=4$ and $c=5$ and the lengths of the measurement vectors are $M=12$. We compare the accuracy of the proposed algorithm with the algorithm in \cite{liu2015remarks} and Zoltowski's method in \cite{zoltowski1994real}. The proposed algorithm uses three coprime channels, while the algorithm in \cite{liu2015remarks} uses two coprime channels and Zoltowski's method uses two undersampled channels with time delay, but these three algorithms are set to use the same number of samples. The ESPRIT algorithm is employed to process the estimate of the autocorrelation matrix, so high-resolution continuous frequency estimates can be obtained. The root mean square error (RMSE) is used to measure the accuracy of the algorithm, which is defined as
\begin{equation}\label{eq12}
  {\rm{RMSE}} = \sqrt {{{\sum\limits_{k = 1}^K {{{\left( {{{\hat f}_k} - {f_k}} \right)}^2}} } \mathord{\left/
 {\vphantom {{\sum\limits_{k = 1}^K {{{\left( {{{\hat f}_k} - {f_k}} \right)}^2}} } K}} \right.
 \kern-\nulldelimiterspace} K}},
\end{equation}
where ${\hat f}_k$ is the estimate of $f_k$. For ease of calculation, the number of frequency components $K$ is assumed to be known. The main steps of the three algorithms can be divided into two parts: the estimation of the autocorrelation matrix and the frequency estimation based on the ESPRIT algorithm. Therefore, the computational complexity of these algorithms is roughly equivalent.

\begin{figure}[!htbp]
\centering
\includegraphics[scale=0.6]{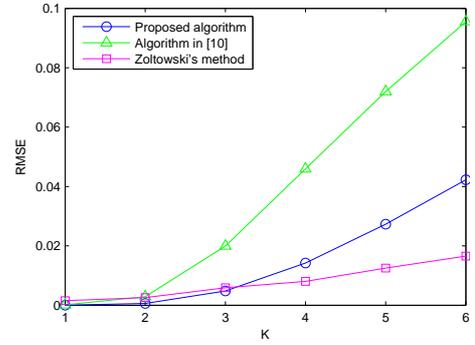}
\caption{ Comparison of RMSEs for different numbers of frequencies. } \label{fig2}
\end{figure}

In the first simulation, the signal-to-noise ratio (SNR) is fixed at 20 dB and the number of frequency components $K$ varies from 1 to 6. Each data point is the average of 10,000 trials. As shown in Fig.~\ref{fig2}, with the increase of the number of frequency components, the RMSEs increase gradually. When $K>3$, Zoltowski's method has the best performance. But when $K\leq 3$, the errors of the proposed algorithm are smaller.

\begin{figure}[!htbp]
\centering
\includegraphics[scale=0.6]{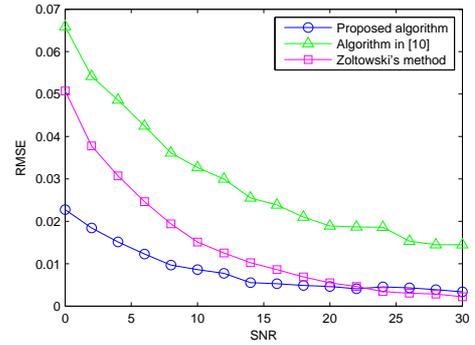}
\caption{ Comparison of RMSEs for different SNRs. } \label{fig3}
\end{figure}

Then we fix the frequency components to $K=3$, and compare the performance of the three algorithms under different SNRs. As shown in Fig.~\ref{fig3}, when ${\text {SNR}}\leq22 \text{dB}$, the proposed algorithm has higher accuracy than the other two algorithms. When ${\text {SNR}}>22 \text{dB}$, the errors of the proposed algorithm are slightly larger than Zoltowski's method. The result indicates that the proposed algorithm has better robustness.


\section{Conclusion}
This letter proposes an algorithm based on subspace techniques to estimate the frequencies of complex sinusoids with multiple-channel coprime sampling. The proposed algorithm is suitable for any number of channels and three coprime channels are used to ensure the resolution of frequency ambiguity. Numerical experiments show that the proposed algorithm has high accuracy and good robustness.
\vskip3pt
\ack{This work is supported by the National Natural Science Foundation of China under Grant 61501335.}

\vskip5pt

\noindent Shan Huang, Haijian Zhang, Hong Sun and Lei Yu (\textit{Signal Processing Laboratory, School of Electronic Information, Wuhan University, China})
\vskip3pt

\noindent E-mail: staronice@whu.edu.cn

\end{document}